\documentclass[aps,preprint]{revtex4}%
\usepackage{amsfonts}
\usepackage{amsmath}
\usepackage{amssymb}
\usepackage{graphicx}%
\begin{document}
\title{Reveal the lost entanglement for accelerated atoms in the high-dimensional
spacetime }
\author{Jiatong Yan$^{1,2}$}
\author{Baocheng Zhang$^1$}
\email{zhangbaocheng@cug.edu.cn}
\author{Qingyu Cai$^{3,4,5}$}
\email{qycai@hainanu.edu.cn}

\affiliation{$^1$School of Mathematics and Physics, China University of Geosciences, Wuhan
430074, China}
\affiliation{$^2$Physics Department, Brown University, Providence, Rhode Island 02912, USA}
\affiliation{$^3$Center for Theoretical Physics, Hainan University, Haikou, 570228, China}
\affiliation{$^4$School of Information and Communication Engineering, Hainan University,
Haikou, 570228, China}
\affiliation{$^5$Peng Huanwu Center for Fundamental Theory, Hefei, 230026, Anhui, China}


\begin{abstract}
When atoms are accelerated in the vacuum, entanglement among atoms will
degrade compared with the initial situation before the acceleration. In this
paper, we propose a novel and interesting view that the lost entanglement can
be recovered completely when the high-dimensional spacetime is exploited, in
the case that the acceleration is not too large, since the entanglement loss
rate caused by the large acceleration is faster than the recovery process. We
also calculate the entanglement change caused by the anti-Unruh effect and
found that the lost entanglement could just be recovered part by the
anti-Unruh effect, and the anti-Unruh effect could only appear for a finite
range of acceleration when interaction time scale is approximately shorter
than the reciprocal of the energy gap in two dimensional spacetime. The limit
case of zero acceleration is also investigated, which gives an analytical
interpretation for the increase or recovery of entanglement.

\end{abstract}
\maketitle


\section{Introduction}

The Unruh effect states that an observer with a uniform acceleration $a$ in
the Minkowski vacuum of a free quantum field would feel a thermal bath of
particles at the temperature $T={\hbar a}/{(2\pi ck_{B})}$ \cite{wgu76}. This
effect was put forward soon after Hawking discovered that a black hole could
emit thermal radiation \cite{swh74} and could help to clarify some conceptual
issues raised by black-hole evaporation. So, the understanding of the Unruh
effect is also significant for Hawking radiation and related problems (i.e.,
information loss problems). In the past years, the Unruh effect was digested
and extended to many different situations (see the review \cite{chm08} and
references therein). A closely related physical phenomenon influenced by the
Unruh effect is quantum entanglement. Quantum entanglement plays a significant
role in quantum information theory and is usually studied in many aspects of
theoretical physics under the background of relativity, such as the black hole
information loss paradox \cite{dh16}, harvesting or extracting information
from the vacuum or thermal background
\cite{harvesting1,harvesting2,harvesting3,harvesting4,harvesting5,harvesting6,harvesting7,harvesting8,harvesting9,harvesting10}%
, quantum metrology \cite{metrology1,metrology2,metrology3}, quantum dynamics
\cite{dynamics1,dynamics2,dynamics3}, and the inner structure of spacetime
\cite{spacetimestructure}.

It has been found widely that acceleration would lead to the decoherence of
quantum states \cite{pt04}, and therefore, lead to the decrease in
entanglement \cite{ast06,mjl10,wj11,ses12,bdl12,ro15} if the initial quantum
state is an entangled state. Sometimes, the acceleration could cause the
increase of entanglement, as given in those phenomena associated with the
anti-Unruh effect \cite{bmm16,gmr16,lzy18,pz20,pz21,zhy21,bm21,chy22,pz23}.
However, no matter what kind of situation, entanglement will become less than
the initial entanglement before the acceleration. This causes a question:
where is the lost entanglement? If the information is carried by entanglement,
where is the information carried by the lost entanglement? Generally, the loss
of entanglement is attributed to the existence of the inaccessible region for
accelerated observers \cite{fm05}. Sometimes, it is also thought that a
similar mechanism to the environment-induced decoherence works for the loss of
entanglement among accelerated atoms \cite{amm95,ky03}. In this paper, we will
investigate the change of entanglement further in the high-dimensional
spacetime \cite{yz22}, and focus on the case in which entanglement between
accelerated atoms can be increased and even recovered with the increasing
spacetime dimension, based on a model of the Unruh-DeWitt detector
\cite{bsd1979} and statistical inversion \cite{Takagi,abg21}.

In this paper, we provide an interesting interpretation that the lost
entanglement is hidden in the high-dimensional spacetime. According to the
superstring theory \cite{jp98}, the dimension of spacetime could be larger
than four. When the particles interact with the vacuum, entanglement among the
particles would be hidden into the compactified extra dimensional spacetime.
Thus, when we measure entanglement in the common 4-dimentional spacetime, a
partial and even the whole entanglement seems to lose. If we extend the
measurement to the higher dimensional spacetime, the initial entanglement
before the interaction can be recovered. In what follows, we will show this
with the example of two atoms in the Minkowski vacuum in the accelerated case
and analyze briefly the entanglement change caused by the (anti-)Unruh effect.
We also study the analytical condition for the anti-Unruh effect and show that
it will not appear in over-2 dimensional spacetime.

\section{Model setup}

We start to discuss the situation that one atom is accelerating in the
$D$-dimensional spacetime. The interaction between the accelerated atom and
the spacetime can be modeled by the interaction Hamiltonian, $H_{I}%
=\lambda\chi\left(  \tau/\sigma\right)  \mu\left(  \tau\right)  \phi\left(
x\left(  \tau\right)  \right)  $, with $\lambda$ the coupling strength. $\tau$
is the atom's proper time along its trajectory $x\left(  \tau\right)
=(t(\tau),\boldsymbol{x}\left(  \tau\right)  )$. $\chi\left(  \tau
/\sigma\right)  $ is a switching function that is used to control the
interaction timescale $\sigma$. $\mu\left(  \tau\right)  $ is the atom's
monopole momentum. $\phi(x(\tau))$ is the scalar field related to the vacuum,
\begin{equation}
\phi(x(\tau))=\int d^{D-1}k\{a_{k}\phi_{k}(x(\tau))+a_{k}^{\dagger}\phi
_{k}^{\ast}(x(\tau))\},
\end{equation}
where $k$ denotes the mode of the scalar field with (Bosonic) annihilation
(creation) operator $a_{k}$ ($a_{k}^{\dagger}$), $a_{k}\left\vert
0\right\rangle =0$ and $a_{k}^{\dag}\left\vert 0\right\rangle =\left\vert
1_{k}\right\rangle $. The superscript $\ast$ denotes the complex conjugate.
$\phi_{k}$ is the Minkowski mode function with the expression as $\phi
_{k}=[2\omega(2\pi)^{D-1}]^{-1/2}e^{ikx(\tau)-\omega_{k}t(\tau)}$ where
$\omega_{k}=|k|$ for the massless vacuum field. Thus, the evolution of the
total quantum state is determined perturbatively by the unitary operator which
up to the first order,
\begin{equation}
U=I+U^{(1)}+O\left(  \lambda^{2}\right)  =I-i\int d\tau H\left(  \tau\right)
+\mathcal{O}\left(  \lambda^{2}\right)  .
\end{equation}
Within the first-order approximation and in the interaction picture, this
evolution is described by \cite{bmm16,lzy18}
\begin{align}
U\left\vert g\right\rangle \left\vert 0\right\rangle  &  =K_{0}\left(
\left\vert g\right\rangle \left\vert 0\right\rangle -i\eta_{_{0}}\left\vert
e\right\rangle \left\vert 1_{k}\right\rangle \right)  ,\nonumber\\
U\left\vert e\right\rangle \left\vert 0\right\rangle  &  =K_{1}\left(
\left\vert e\right\rangle \left\vert 0\right\rangle -i\eta_{_{1}}\left\vert
g\right\rangle \left\vert 1_{k}\right\rangle \right)  , \label{foe}%
\end{align}
where $K_{0,1}$ is the state normalization factor. It is noted that the
created state $\left\vert 1_{k}\right\rangle $ is dependent on the wave vector
$k$, so the coupling in Eq. (\ref{foe}) has to be understood by writing
$\eta_{_{0}}\left\vert 1\right\rangle =\lambda\int d^{D-1}kI_{+,k}\left\vert
1_{k}\right\rangle $ and $\eta_{_{1}}\left\vert 1\right\rangle =\lambda\int
d^{D-1}kI_{-,k}\left\vert 1_{k}\right\rangle $ where $I_{\pm,k}$ is given as
\begin{equation}
I_{\pm,k}=\frac{1}{\sqrt{2(2\pi)^{D-1}\omega}}\int_{-\infty}^{\infty}%
\chi\left(  \tau/\sigma\right)  \exp[\pm i\Omega\tau+i\omega t\left(
\tau\right)  -ikx\left(  \tau\right)  ]{d\tau}\text{.}%
\end{equation}
The notations $\eta_{_{0}}\left\vert 1\right\rangle $ and $\eta_{_{1}%
}\left\vert 1\right\rangle $ is inseparable, and $\eta_{_{0}}$ and $\eta
_{_{1}}$ are related to the excitation and deexcitation probability of the
atom, which are gotten from
\begin{align}
tr_{\phi}(|\eta_{0}^{2}||1_{k}\rangle\langle1_{k}|)  &  =P_{+},\nonumber\\
tr_{\phi}(|\eta_{1}^{2}||1_{k}\rangle\langle1_{k}|)  &  =P_{-}, \label{P121}%
\end{align}
where $tr_{\phi}$ denotes tracing out the field degrees of freedom. $P_{+}$
and $P_{-}$ are excitation and deexcitation probabilities of the atom,
respectively. Rewriting the integral about the field mode $k$ in expressions
for the probabilities in the manner of Wightman function $W(\tau-\tau^{\prime
})$, we obtain
\begin{equation}
P_{\pm}=\int\int d\tau d\tau^{\prime}W(\tau-\tau^{\prime})\chi(\tau
/\sigma)\chi(\tau^{\prime}/\sigma)e^{\mp i\Omega(\tau-\tau^{\prime})},
\label{PPP}%
\end{equation}%
\begin{equation}
X_{\pm}=\int\int d\tau d\tau^{\prime}W(\tau-\tau^{\prime})\chi(\tau
/\sigma)\chi(\tau^{\prime}/\sigma)e^{\pm i\Omega(\tau+\tau^{\prime})},
\label{XXX}%
\end{equation}
where the relation $W(\tau-\tau^{\prime})=\langle0|\phi(x(\tau))\phi
(x(\tau^{\prime}))|0\rangle$ is assured by the time translational invariance
property of the Wightman function in the stationary spacetime, $X_{+}=\eta
_{0}\eta_{1}^{\ast}$, $X_{-}=\eta_{0}^{\ast}\eta_{1}$. The switching function
$\chi(\tau)$ is chosen as the Gaussian function $e^{-\tau^{2}/2\sigma^{2}}$.

Now, we consider two two-level atoms as detectors with the initial state as
$|\Psi_{i}\rangle=\alpha|g\rangle|e\rangle+\beta|e\rangle|g\rangle$, where the
complex coefficients satisfies $\left\vert \alpha\right\vert ^{2}+\left\vert
\beta\right\vert ^{2}=1$, and $|g\rangle$ ($|e\rangle$) is the ground
(excited) state of the atom. The initial joint state of detector-field is
rewritten as $|\Psi_{i}\rangle|0\rangle|0\rangle$ where $|0\rangle$ denotes
the vacuum field. After interaction with the vacuum, the joint state becomes
\begin{align}
&  |\Psi_{f}\rangle=C_{0}C_{1}[\left(  \alpha\left\vert g\right\rangle
\left\vert e\right\rangle +\beta\left\vert e\right\rangle \left\vert
g\right\rangle \right)  \left\vert 0\right\rangle \left\vert 0\right\rangle
\nonumber\\
&  -i\left(  \alpha\eta_{_{1}}\left\vert g\right\rangle \left\vert
g\right\rangle +\beta\eta_{_{0}}\left\vert e\right\rangle \left\vert
e\right\rangle \right)  \left\vert 0\right\rangle \left\vert 1_{k}%
\right\rangle \nonumber\\
&  -i\left(  \beta\eta_{_{1}}\left\vert g\right\rangle \left\vert
g\right\rangle +\alpha\eta_{_{0}}\left\vert e\right\rangle \left\vert
e\right\rangle \right)  \left\vert 1_{k}\right\rangle \left\vert
0\right\rangle \nonumber\\
&  +\left(  \alpha\eta_{_{0}}\eta_{_{1}}\left\vert e\right\rangle \left\vert
g\right\rangle +\beta\eta_{_{0}}\eta_{_{1}}\left\vert g\right\rangle
\left\vert e\right\rangle \right)  \left\vert 1_{k}\right\rangle \left\vert
1_{k}\right\rangle ]. \label{js}%
\end{align}
We can trace out the field state in Eq. (\ref{js}) to obtain the state of the
two atoms. The density matrix of the final two-atom state is
\begin{equation}
\rho=\mathcal{K}\left[  {%
\begin{array}
[c]{cccc}%
P_{-} & 0 & 0 & 2\alpha\beta X_{+}\\
0 & \alpha^{2}+\beta^{2}P_{+}P_{-} & \alpha\beta+\alpha\beta P_{+}P_{-} & 0\\
0 & \alpha\beta+\alpha\beta P_{+}P_{-} & \alpha^{2}+\beta^{2}P_{+}P_{-} & 0\\
2\alpha\beta X_{-} & 0 & 0 & P_{+}%
\end{array}
}\right]  \label{dm}%
\end{equation}
where $\mathcal{K}$ is the normalization factor $1/(1+P_{+}+P_{-}+P_{+}P_{-})$
which assures the sum of the probabilities of the system in each state is $1$.

In the general scenario we consider, the initial two-atom state is a pure one
in flat spacetime. When the two atoms interact with the vacuum, the bipartite
state becomes mixed, and entanglement between the two atoms is changed. The
change of entanglement can be quantified by concurrence \cite{wkw98} which is
a widely used entanglement measure for bipartite mixed state. Concurrence is
defined by
\begin{equation}
\mathcal{C}\left(  \rho\right)  =\max\{0,\lambda_{1}-\lambda_{2}-\lambda
_{3}-\lambda_{4}\},
\end{equation}
where $\lambda_{1}$, $\lambda_{2}$, $\lambda_{3}$, $\lambda_{4}$ are the
eigenvalues of the Hermitian matrix $\sqrt{\sqrt{\rho}\widetilde{\rho}%
\sqrt{\rho}}$ with $\widetilde{\rho}=\left(  \sigma_{y}\otimes\sigma
_{y}\right)  \rho^{\ast}\left(  \sigma_{y}\otimes\sigma_{y}\right)  $ the
spin-flipped state of $\rho$, $\sigma_{y}$ being the $y$-component Pauli
matrix, and the eigenvalues listed in decreasing order. Because the density
matrix in Eq. (\ref{dm}) always remain the X form, the four eigenvalues are%
\begin{align}
&  \mathcal{K}(\sqrt{P_{+}P_{-}}+2\alpha\beta|X|),\nonumber\\
&  \mathcal{K}(\sqrt{\alpha^{2}\beta^{2}(1+P_{+}^{2}P_{-}^{2})+(\alpha
^{4}+\beta^{4})P_{+}P_{-}}+\alpha\beta(1+P_{+}P_{-})),\nonumber\\
&  \mathcal{K}(\sqrt{P_{+}P_{-}}-2\alpha\beta|X|),\nonumber\\
&  \mathcal{K}(\sqrt{\alpha^{2}\beta^{2}(1+P_{+}^{2}P_{-}^{2})+(\alpha
^{4}+\beta^{4})P_{+}P_{-}}-\alpha\beta(1+P_{+}P_{-})).
\end{align}
Then the concurrence could be written as
\begin{align}
\mathcal{C}(\rho)  &  =\mathcal{K}\max\{0,\nonumber\\
&  4\alpha\beta|X|-2\sqrt{(\alpha^{4}+\beta^{4})P_{+}P_{-}+\alpha^{2}\beta
^{2}(1+(P_{+}P_{-})^{2})}\nonumber\\
&  -2\alpha\beta(1+P_{+}P_{-})-2\sqrt{P_{+}P_{-}}\}.
\end{align}
For a further calculation, the Wightman function should be given to calculate
$P_{\pm}$ in Eq. (\ref{PPP}) and $X_{\pm}$ in Eq. (\ref{XXX}).

\section{Entanglement recovery}

In this section, we will consider two uniformly accelerated detectors in the
Minkowski vacuum. Their trajectories are defined as
\begin{equation}
t(\tau)=\frac{1}{a}\sinh{a\tau},x_{0}(\tau)=\frac{1}{a}\cosh{a\tau}%
,x_{1}=x_{2}=\cdots=x_{D-2}=0. \label{at}%
\end{equation}
Along the trajectory of the uniformly accelerated detector, the Wightman
function is given as \cite{Takagi}
\begin{equation}
W_{a}(x,x^{\prime})=\mathcal{C}_{0}(\frac{a}{2i})^{D-2}(\sinh[\frac
{a(\Delta\tau-i\epsilon)}{2}])^{-(D-2)}, \label{wa}%
\end{equation}
where the subscript $a$ denotes the accelerated case.

As the prediction of the Unruh effect, the background temperature felt by a
uniformly accelerated is proportional to the acceleration, so a UDW detector
will click more often with the acceleration increasing, but in some specific
conditions, a UDW detector will click less often when the acceleration is
increasing, i.e. the detector will feel a lower temperature. This
counterintuitive phenomenon is dubbed as the anti-Unruh effect. With the
Wightman function, the transition probability can be calculated.

\begin{figure}[ptb]
\centering
\includegraphics[width=0.7\columnwidth]{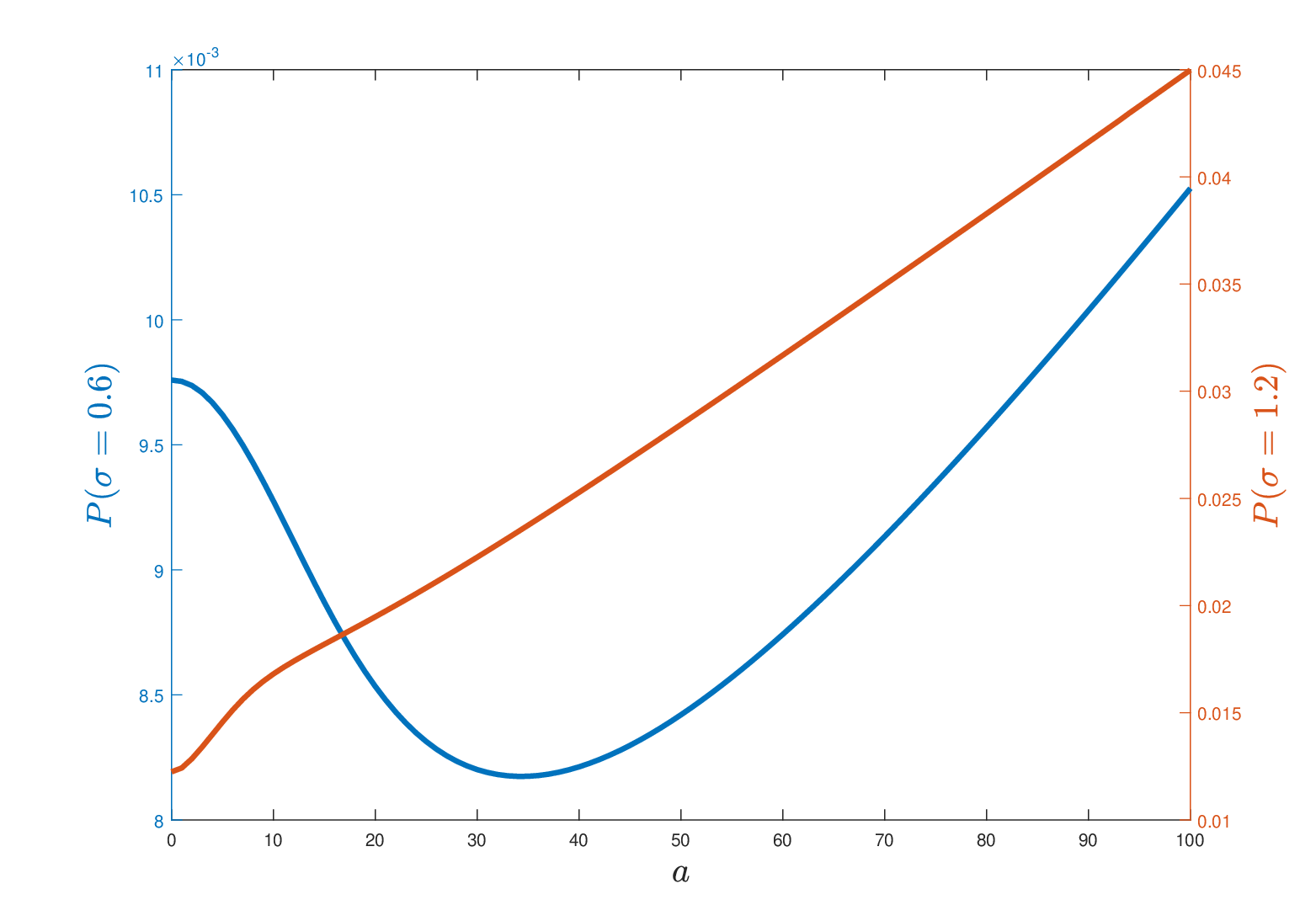}\caption{(Color online) The
transition probability as a function of the acceleration $a$. The red and blue
lines stand for the Unruh ($\sigma=0.6$) and anti-Unruh ($\sigma=1.2$)
effects, respectively. Other parameters are taken as $\lambda=0.05$,
$\omega=1$, $\Lambda=10^{-7}$.}%
\label{Fig1}%
\end{figure}

Typically, the thermal noise produced by the Unruh effect will lead to the
decoherence of a quantum state and the decrease of entanglement, so it is
natural to think that whether the `colding' phenomenon caused by the
anti-Unruh effect will lead to the complete recovery of the loss entanglement
between two atoms caused by the interaction with the vacuum? Before analyzing
this, we first interpret in what cases the anti-Unruh effect can appear.

\subsection{Anti-Unruh effect}

In two-dimensional spacetime, the Wightman function for the accelerated atom
is \cite{Takagi}
\begin{equation}
W_{2a}=-\frac{1}{2\pi}Log[\frac{2i}{a}\sinh(\frac{at}{2})], \label{2dwa}%
\end{equation}
which cannot be obtained directly from Eq. (\ref{wa}), and requires a
dimensional regulation \cite{bd1978}. When the acceleration is small, using
the expansion $Log[\frac{2i}{a}\sinh(\frac{a\tau}{2})]=Log[it]+\frac{a^{2}%
\tau^{2}}{24}+\mathcal{O}(a^{4})$, we get
\begin{equation}
W_{2a}=-\frac{1}{2\pi}(Log[i\tau]+\frac{a^{2}\tau^{2}}{24})+\mathcal{O}%
(a^{4}). \label{ser1}%
\end{equation}
The first term in this equation is noting but the two-dimensional Wightman
function of a static detector ($a=0$).

Now we analyze the second term of Eq. (\ref{ser1}) to deduce the exact
condition of the anti-Unruh effect. Inserting Eq. (\ref{ser1}) into Eq.
(\ref{PPP}), one obtain
\begin{equation}
P_{2a}=P_{2s}+\frac{1}{12}\pi^{2}a^{2}\sigma^{4}e^{-\sigma^{2}\Omega^{2}%
}(1-2\Omega^{2}\sigma^{2})+\mathcal{O}(a^{4}).
\end{equation}
When $1-2\Omega^{2}\sigma^{2}<0$, the second term is negative and leads to the
anti-Unruh effect. So the exact condition for the anti-Unruh effect is
\begin{equation}
\sigma\Omega<\frac{1}{\sqrt{2}},
\end{equation}
which is consistent with the result in Ref. \cite{wys2023} where it was also
shown that the condition for anti-Unruh effect is dependent on the concrete
form of switching function.

We can also just go directly from the mode expansion, combined with the
accelerated trajectory (\ref{at}), and get the transition probability
\begin{align}
P_{2a}  &  =\int_{-\infty}^{\infty}\frac{dk}{4\pi|k|}|\int d\tau
e^{i\Omega\tau+i|k|t(\tau)-ikx(\tau)-\tau^{2}/2\sigma^{2}2}|^{2}\nonumber\\
&  =-\int_{-\infty}^{-\Lambda}\frac{dk}{4\pi k}\int d\tau d\tau^{\prime
}e^{i\Omega(\tau-\tau^{\prime})}e^{-\frac{t^{2}+t^{\prime2}}{2\sigma^{2}}%
}e^{-i\frac{k}{a}(e^{a\tau}-e^{a\tau^{\prime}})}\nonumber\\
&  +\int_{\Lambda}^{\infty}\int d\tau\int d\tau^{\prime}e^{i\Omega(\tau
-\tau^{\prime})}e^{-\frac{\tau^{2}+\tau^{\prime2}}{2\sigma^{2}}}e^{-i\frac
{k}{a}(e^{-a\tau}-e^{-a\tau^{\prime}})},
\end{align}
where we take the cutoff at points $k=\pm\Lambda$ to avoid the divergent
problem. After performing the $k$ integral, we obtain
\begin{equation}
P_{2a}=\frac{1}{4\pi}\int d\tau d\tau^{\prime}e^{i\Omega(\tau-\tau^{\prime}%
)}e^{-\tau^{2}/2\sigma^{2}}e^{-\tau^{\prime2}/2\sigma^{2}}[\Gamma
(0,\frac{i\Lambda}{a}(e^{-a\tau}-e^{-a\tau^{\prime}}))+\Gamma(0,-\frac
{i\Lambda}{a}(e^{a\tau}-e^{a\tau^{\prime}}))]. \label{P2D}%
\end{equation}
where $\Gamma$ stands for the incomplete Gamma function.

Similarly, the cross term is
\begin{equation}
X_{2a}=\frac{1}{4\pi}\int d\tau d\tau^{\prime}e^{i\Omega(\tau+\tau^{\prime}%
)}e^{-\tau^{2}/2\sigma^{2}}e^{-\tau^{\prime2}/2\sigma^{2}}[\Gamma
(0,\frac{i\Lambda}{a}(e^{-a\tau}-e^{-a\tau^{\prime}}))+\Gamma(0,-\frac
{i\Lambda}{a}(e^{a\tau}-e^{a\tau^{\prime}}))]. \label{X2D}%
\end{equation}

These are the numerical integrals we need to perform, and they can be
simplified by the Taylor expansion for the case of $x<1$, $\Gamma
(0,x)=-\gamma-Log[x]+\mathcal{O}[x]$, where $\gamma$ is the Euler constant.
Applying $P_{2a}$ and $X_{2a}$ to Eq. (\ref{P2D}), we get
\begin{equation}
P_{2a}\simeq\frac{1}{4\pi}\int d\tau d\tau^{\prime}e^{i\Omega(\tau
-\tau^{\prime})}e^{-\tau^{2}/2\sigma^{2}}e^{-\tau^{\prime2}/2\sigma^{2}%
}(-2\gamma-Log[\frac{i\Lambda}{a}(e^{-a\tau}-e^{-a\tau^{\prime}}%
)]-Log[-\frac{i\Lambda}{a}(e^{at}-e^{at^{\prime}})]), \label{P2Da}%
\end{equation}
and the cross term
\begin{equation}
X_{2a}\simeq\frac{1}{4\pi}\int d\tau d\tau^{\prime}e^{i\Omega(\tau
+\tau^{\prime})}e^{-\tau^{2}/2\sigma^{2}}e^{-\tau^{\prime2}/2\sigma^{2}%
}(-2\gamma-Log[\frac{i\Lambda}{a}(e^{-a\tau}-e^{-a\tau^{\prime}}%
)]-Log[-\frac{i\Lambda}{a}(e^{at}-e^{at^{\prime}})]). \label{x2da}%
\end{equation}
Depending on the cut-off point $\Lambda$, the results will be different, so
our calculation of the two-dimensional case is only a qualitative analysis of
the entanglement change. The integral range of $\tau$ is chosen as
$[-10\sigma,10\sigma]$ and we can suppress the error induced by integral range
at $e^{-100}\sim10^{-43}$.

Fig. 1 shows the Unruh and anti-Unruh effects for a uniformly accelerated
detector in two dimensional spacetime. It is seen from the blue line of Fig. 1
that the anti-Unruh effect can only maintain for a finite range of
acceleration. After a range of decrease, the transition probability of a
detector will increase with the acceleration increasing. That is because that
the leading order term of Eq. (\ref{wa}) decreases with increasing
acceleration, which leads to the anti-Unruh effect, but the higher order terms
also increases with increasing acceleration. When acceleration is small, the
higher order terms could be neglected and the leading-order term works and
induces the anti-Unruh effect; when acceleration is large, the higher order
terms are dominant and suppress the decrease caused by the leading-term, so in
this case, no anti-Unruh effect appears.

For the over-two-dimensional case, inserting Eq. (\ref{wa}) into Eq.
(\ref{PPP}), and making some simplifications, we get the transition
probability
\begin{equation}
P_{a}=\sqrt{\pi}\sigma\mathcal{C}(a/2i)^{D-2}(\sinh[(a(\Delta\tau
-i\epsilon))/2])^{-(D-2)}, \label{Pa}%
\end{equation}
and substituting the expansion
\begin{equation}
(\frac{a}{2i})^{D-2}(\sinh[(a(\Delta\tau-i\epsilon))/2])^{-(D-2)}=\frac
{1}{(\Delta\tau-i\epsilon)^{D-2}}-(D-2)\frac{a^{2}}{24}\frac{1}{(\Delta
\tau-i\epsilon)^{D-4}}+\mathcal{O}(a^{4})) \label{sh}%
\end{equation}
into Eq. (\ref{Pa}), we can obtain
\begin{equation}
P_{a}=\sqrt{\pi}\sigma\mathcal{C}\int d\tau e^{-\tau^{2}/2\sigma^{2}%
}e^{-i\Omega\tau}((-1)^{-\frac{D-2}{2}}\frac{1}{(\tau-i\epsilon)^{D-2}%
}+(-1)^{-\frac{D-4}{2}}\frac{a^{2}}{4}\frac{1}{(\tau-i\epsilon)^{D-4}%
}+\mathcal{O}(a^{4})), \label{Pae}%
\end{equation}
where the first term is just exactly the transition probability of a static
detector in the $D$-dimensional spacetime (see the appendix for the details),
and the second term is proportional to the square of the acceleration $a$. The
two terms are both positive, so the transition probability will increase with
the increasing acceleration, which means that no anti-Unruh effect appears.
The higher-order terms beyond $a^{2}$ are also proportional to a power of
acceleration, but when $a$ is smaller than $1$, these terms are small and
could be neglected when analyzing the trend of transition probability. Thus,
we extend the conclusion of Ref. \cite{wys2023} that no anti-Unruh effect
exists from (3+1) dimensional spacetime to higher dimensional spacetime.

The anti-Unruh effect can only appear in two dimensional spacetime, and the
interaction timescale $\sigma$ should be small ($\sigma\Omega<1/\sqrt{2}$) to
induce the anti-Unruh effect. Under this condition, we calculate the
concurrence, as is shown in Fig. 2. It is seen that the anti-Unruh effect can
lead to the increase of entanglement, but it cannot recover the initial entanglement.

\begin{figure}[ptb]
\centering
\includegraphics[width=0.7\columnwidth]{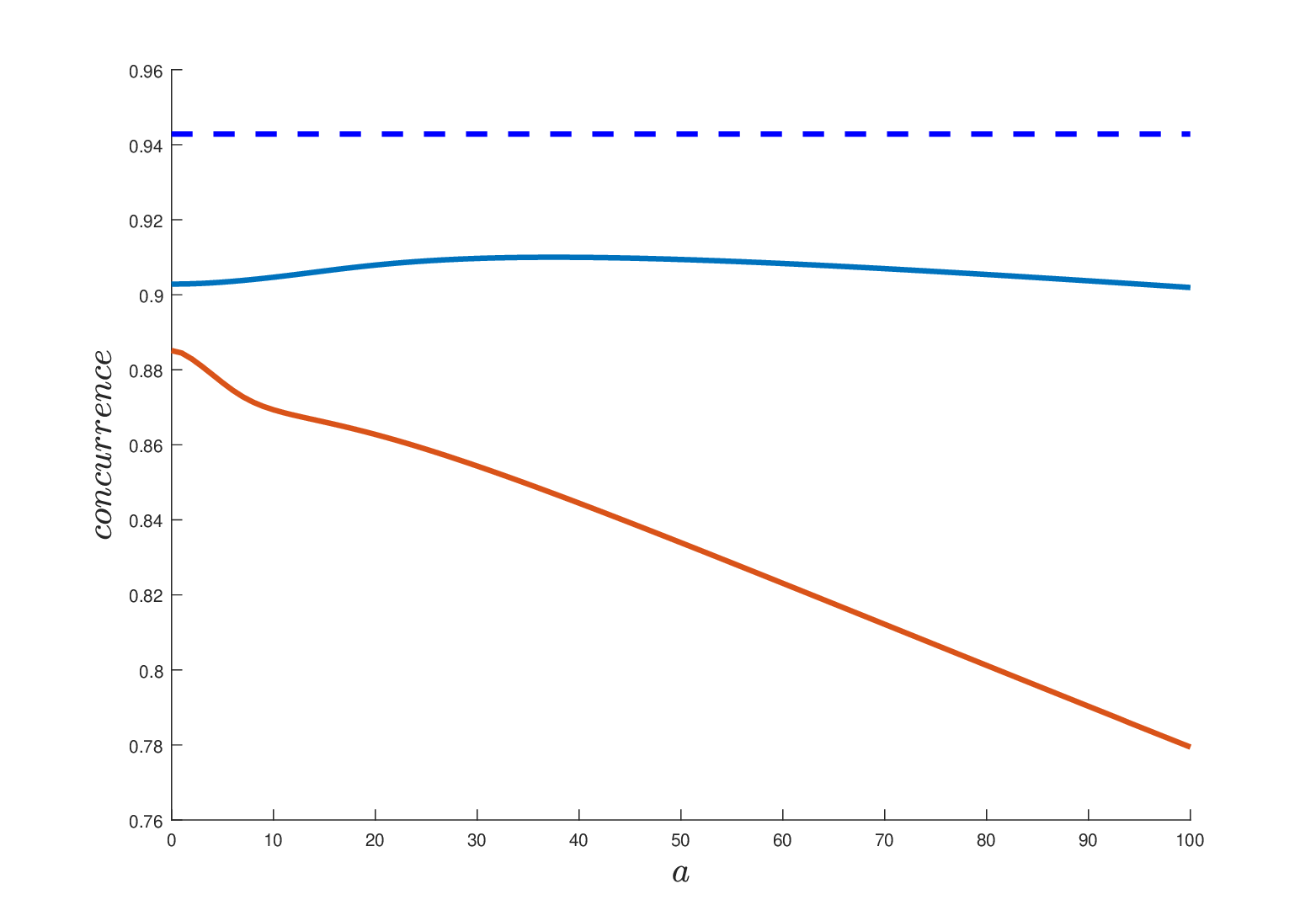} \caption{(Color online) The
concurrence as a function of the acceleration $a$. The solid red and blue
lines represents the Unruh and anti-Unruh effects, respectively. The dashed
blue line denotes the quantity of the initial entanglement. Here the parameter
is taken as $\alpha= 1/\sqrt{3}$, and other parameters are the same as in
Fig.1.}%
\label{Fig2}%
\end{figure}

\subsection{General accelerated case}

In the accelerated case, the expressions for $P$ and $X$ are
\begin{equation}
P_{\pm a}=Ma^{-3+D}\int dte^{\mp\frac{2i\Omega t}{a}-\frac{t^{2}}{a^{2}%
\sigma^{2}}}\frac{\sinh{(t-i\epsilon)}^{D-2}-(t-i\epsilon)^{D-2}}%
{(t-i\epsilon)^{D-2}\sinh{(t-i\epsilon)}^{D-2}}+P_{\pm s} \label{Paa}%
\end{equation}
and
\begin{equation}
X_{a}=Ma^{-3+D}e^{-\sigma^{2}\Omega^{2}}\int dte^{-\frac{t^{2}}{a^{2}%
\sigma^{2}}}\frac{\sinh{(t-i\epsilon)}^{D-2}-(t-i\epsilon)^{D-2}}%
{(t-i\epsilon)^{D-2}\sinh{(t-i\epsilon)}^{D-2}}+X_{s} \label{Xaa}%
\end{equation}
where the constants $M=i^{-D}2^{1-D}\pi^{\frac{1-D}{2}}\sigma\Gamma(\frac
{D}{2})$, and the detailed calculation is given in the appendix.

\begin{figure}[ptb]
\centering
\includegraphics[width=0.7\columnwidth]{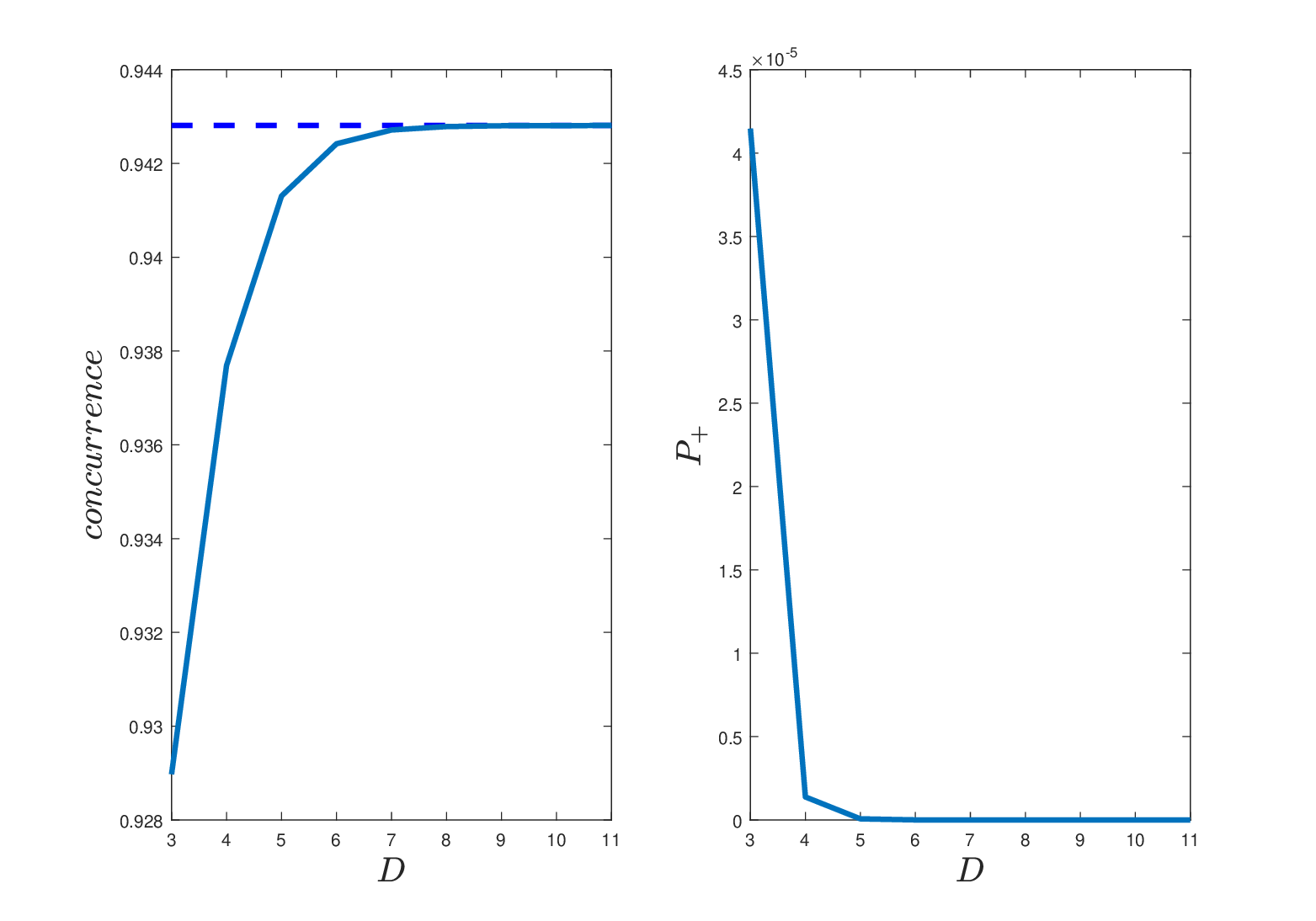} \caption{(Color online) The
left panel shows concurrence of two uniformly accelerated detectors in the
`small acceleration' case with respect to the dimension $D$ of the spacetime.
The right panel shows the corresponding transition probability from the ground
state to the excited state. Parameters are taken as $a = 0.2$, $\lambda= 0.1$,
$\sigma= 2$, $\Omega= 1$, and $\alpha= 1/\sqrt{3}$.}%
\label{Fig3}%
\end{figure}

\begin{figure}[ptb]
\centering
\includegraphics[width=0.7\columnwidth]{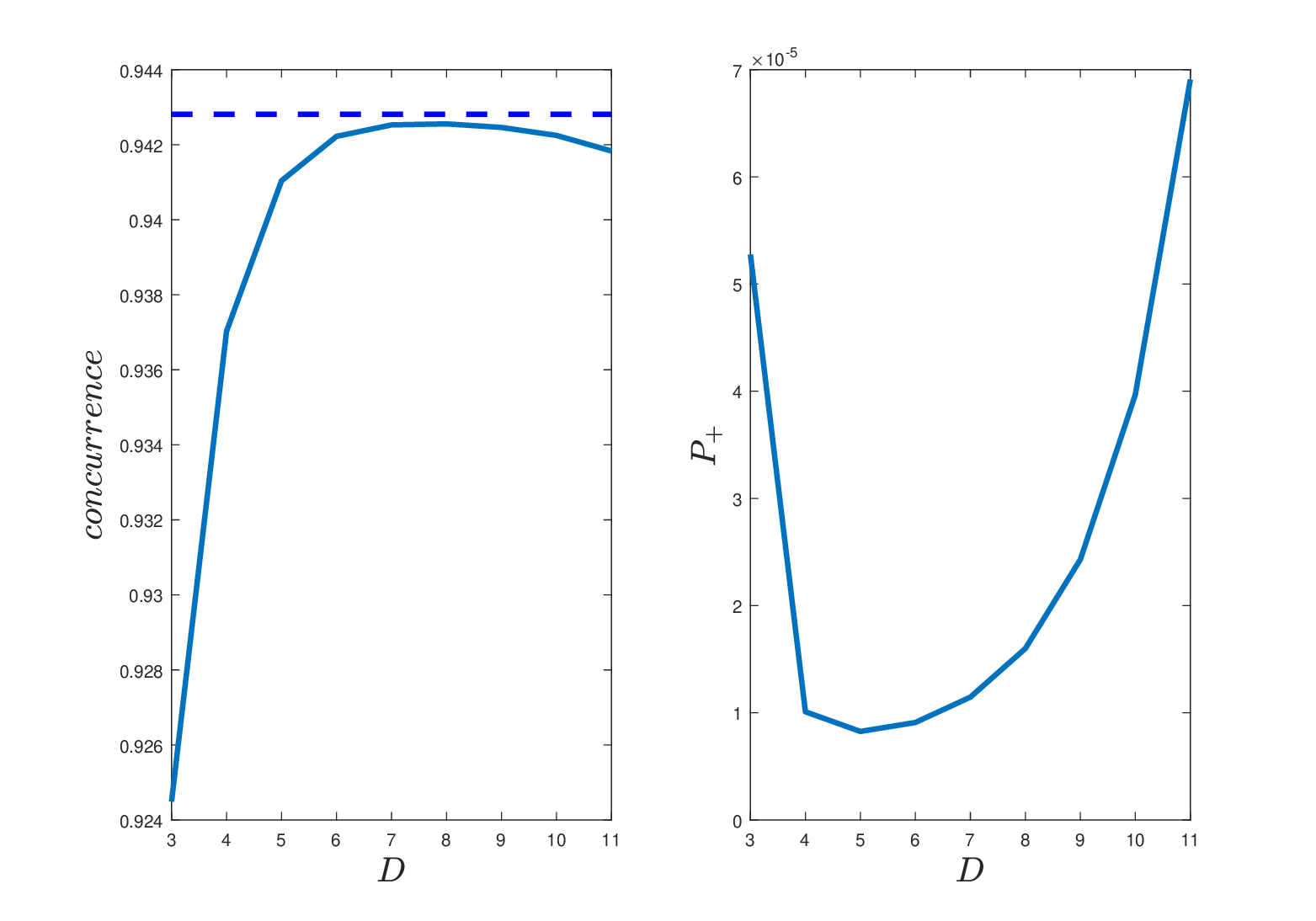} \caption{(Color online) The
left panel shows the concurrence of two uniformly accelerated detectors in the
`large acceleration' case with respect to the dimension $D$ of the spacetime.
The right panel shows the corresponding transition probability from the ground
state to the excited state. $a = 3$, and other parameters are the same as in
Fig.3.}%
\label{Fig4}%
\end{figure}

The left panel in the Fig. 3 presents the change of entanglement in the
general accelerated case. It is amazing that the entanglement is recovered
almost completely in higher spacetime dimensions. The range of abscissa $D$ is
from $3$ to $11$ which is the prediction of the spacetime dimension in
super-string theory. The reason is given in the right panel of the Fig. 3, in
which the transition probability approaches to zero when the spacetime
dimension increases. This shows that the behavior of accelerated atoms in the
high dimensional spacetime looks as if they are remained still in the
4-dimensional Minkowski vacuum.

However, the entanglement recovery occurs only in the situation that the
acceleration is small. And the entanglement cannot be recovered when the
acceleration is large, because the Unruh thermal bath will lead to more
entanglement loss as the spacetime dimension increases, which covers the
entanglement recovery process. We show this case in Fig. 4. From the
mathematical view, the higher-order terms in Eq. (\ref{Pae}) are all
proportional to a specific power of $a$, so when $a$ is small ($a\lesssim c$,
$c$ is the speed of light which does not appear in the expressions since
natural units are used in this paper), the influence of these terms on the
transition probability is small and then the acceleration will not spoil the
recovery of entanglement; when $a$ is large ($a\gtrsim c$), these terms are
also quite large and then the value of transition probability will be
influenced significantly, the entanglement recovery process is spoiled.

But what if we extend the spacetime dimensionality $D$ to a larger value? In
fact, as long as the detectors are accelerated, the entanglement between two
detectors will turn to loss after a period of recovery (if the acceleration is
large enough, the entanglement will degrade directly). The larger the
acceleration is, the longer the recovery process will be. We show this
phenomenon in Fig. 5.

\begin{figure}[ptb]
\centering
\includegraphics[width=1\columnwidth]{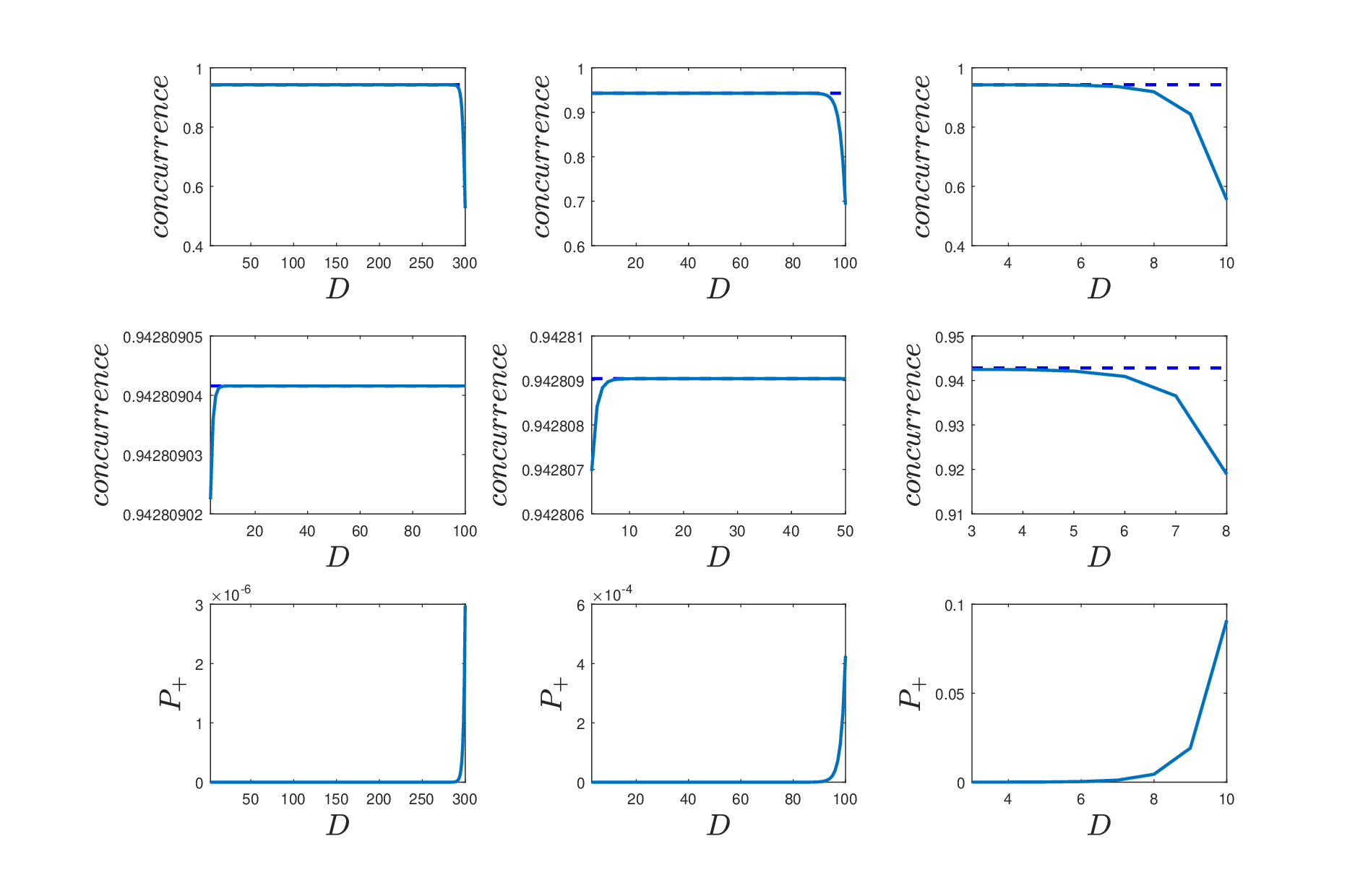}\caption{(Color online) The
concurrence of two uniformly accelerated detectors when spacetime dimension
$D$ extends to a larger value. Parameters are $\sigma=2$, $\Omega=1$,
$\alpha=1/\sqrt{3}$. The values of acceleration $a$ for three columns are
$a=0.5$, $a=0.94$ and $a=10$ respectively from left to right. The value of
coupling constant $\lambda$ for the three columns are $\lambda=10^{-4}$,
$\lambda=10^{-3}$ and $\lambda=10^{-2}$ respectively from left to right. The
second row shows a corresponding zoomed version of the first row. The third
row shows the transition probabilities from the ground state to the excited
state, corresponding to the cases presented in the first row.}%
\label{Fig5}%
\end{figure}

\subsection{The limit case of zero acceleration}

It is noted that the transition probability is not zero for the case of $a=0$
in Eq. (\ref{Paa}), which is the result of the finite interaction time or the
switching function \cite{ls2008,bmm2013,mhm2022}, different from the zero
transition probability in the flat spacetime for the infinite interaction
time. In this subsection, we study the entanglement recovery in the limit case
of zero acceleration and explore the condition for the recovery analytically.

The pullback of the Wightman function of a static detector in the Minkowski
spacetime is written as \cite{Takagi}
\begin{equation}
W(t,t^{\prime})=\mathcal{C}_{0}(-1)^{-\frac{D-2}{2}}\frac{1}{[(t-t^{\prime
}-i\epsilon)^{2}-|{x}-{x^{\prime}}|^{2}]^{\frac{D-2}{2}}}. \label{ws}%
\end{equation}
where $\epsilon\rightarrow0^{+}$, $\mathcal{C}_{0}=(4\pi^{D/2})^{-1}%
\Gamma(\frac{D-2}{2})$ where $\Gamma$ stands for Gamma function, and
$t,x\equiv(t(\tau),x_{0}(\tau),\cdots,x_{D-2}(\tau))$. After a simple but
tedious calculation given in the appendix, we can get the expressions of
$P_{\pm s}$ and $X_{\pm s}$,%

\begin{equation}
P_{\pm s}=%
\begin{cases}
\mathcal{C}_{1}\Omega_{1}F_{1}[2-D/2,3/2,-\sigma^{2}\Omega^{2}]-\mathcal{C}%
_{2}f^{(D-3)}(0),D\ is\ odd\\
\mathcal{C}_{31}F_{1}[\frac{3-D}{2},\frac{1}{2},-\sigma^{2}\Omega
^{2}]-\mathcal{C}_{4}f^{(D-3)}(0),D\ is\ even
\end{cases}
\label{Ps}%
\end{equation}
where
\begin{equation}
X_{s}=e^{-\Omega^{2}\sigma^{2}}%
\begin{cases}
-\mathcal{C}_{4}g^{(D-3)}(0),D\ is\ odd\\
\mathcal{C}_{3}\sigma^{4-D}\pi^{-\frac{1}{2}}-\mathcal{C}_{4}g^{(D-3)}%
(0),D\ is\ odd
\end{cases}
\label{Xs}%
\end{equation}
where the subscribe $s$ denotes the static case, $\mathcal{C}_{1}=2^{2-D}%
\pi^{\frac{3-D}{2}}\sigma^{5-D}$, $\mathcal{C}_{2}=\frac{i^{3-D}\pi
^{1-\frac{D}{2}}\Gamma(\frac{D-2}{2})}{4(D-3)!}$, $\mathcal{C}_{3}%
=(2i)^{-D}\pi^{\frac{1-D}{2}}(D-1)\sigma^{4-D}\Gamma(\frac{1-D}{2}%
)\Gamma(\frac{D-2}{2})$, $\mathcal{C}_{4}=\frac{i^{3-D}\pi^{1-\frac{D}{2}%
}\Gamma(\frac{D-2}{2})}{4(D-3)!}$, $f(x)=e^{-x^{2}/4\sigma^{2}\mp i\Omega x}$,
and $g(x)=e^{-x^{2}/4\sigma^{2}}$. For the $X$ expression, it is shown that
$X_{+s}=X_{-s}\equiv X_{s}$.

It is seen that from Fig. 6 that the entanglement is recovered almost
completely in higher spacetime dimensions in this case, as expected. It is
also found that the entanglement recovery rate is larger compared with the
accelerated case. But it should be noted that only when the interaction
timescale is relatively long compared with the energy gap $\Omega$,
entanglement could be recovered almost completely. That is because that the
structure of higher dimensional spacetime is more complicated, so it will take
longer time to recover the lost entanglement. When the interaction timescale
is relatively short compared with the energy gap $\Omega$, with the spacetime
dimensionality increasing, the entanglement between two atoms will first
recover partly, and then degrade because the interaction time is not long
enough to recover entanglement. We show this case in Fig. 7. When the
interaction time is very short, the entanglement will degrade directly with
spacetime dimension increasing. This is easily understandable since the atoms
has not interacted sufficiently with the vacuum.

To give a concrete condition of entanglement recovery, we denote the first
term in Eq. (\ref{Ps}) as $P_{1}$ and set spacetime dimensionality $D$ as an
even number, then let $P_{1}(D)$ divide $P_{1}(D-2)$, and we get
\begin{equation}
\frac{P_{1}(D)}{P_{1}(D-2)}=\frac{(D-4)}{4\sigma^{2}\pi(D-3)}\frac{_{1}%
F_{1}[\frac{3-n}{2},\frac{1}{2},-\sigma^{2}\Omega^{2}]}{_{1}F_{1}[\frac
{5-n}{2},\frac{1}{2},-\sigma^{2}\Omega^{2}]}.
\end{equation}
When $D$ is very large, this equation could be simplified further
\begin{equation}
\frac{P_{1}(D)}{P_{1}(D-2)}=4\pi\sigma^{2}. \label{DeltaP1}%
\end{equation}
Thus, when $4\pi\sigma^{2}<1$, i.e. $\sigma<\frac{1}{2\sqrt{\pi}}$, $P_{1}$
decreases with $D$ increasing, and the lost entanglement could be recovered.
Because the second term in Eq. (\ref{Ps}) and $P_{1}$ shares the same trend
with increasing $D$, we only need to analyze the first term $P_{1}$.

\begin{figure}[ptb]
\centering
\includegraphics[width=0.7\columnwidth]{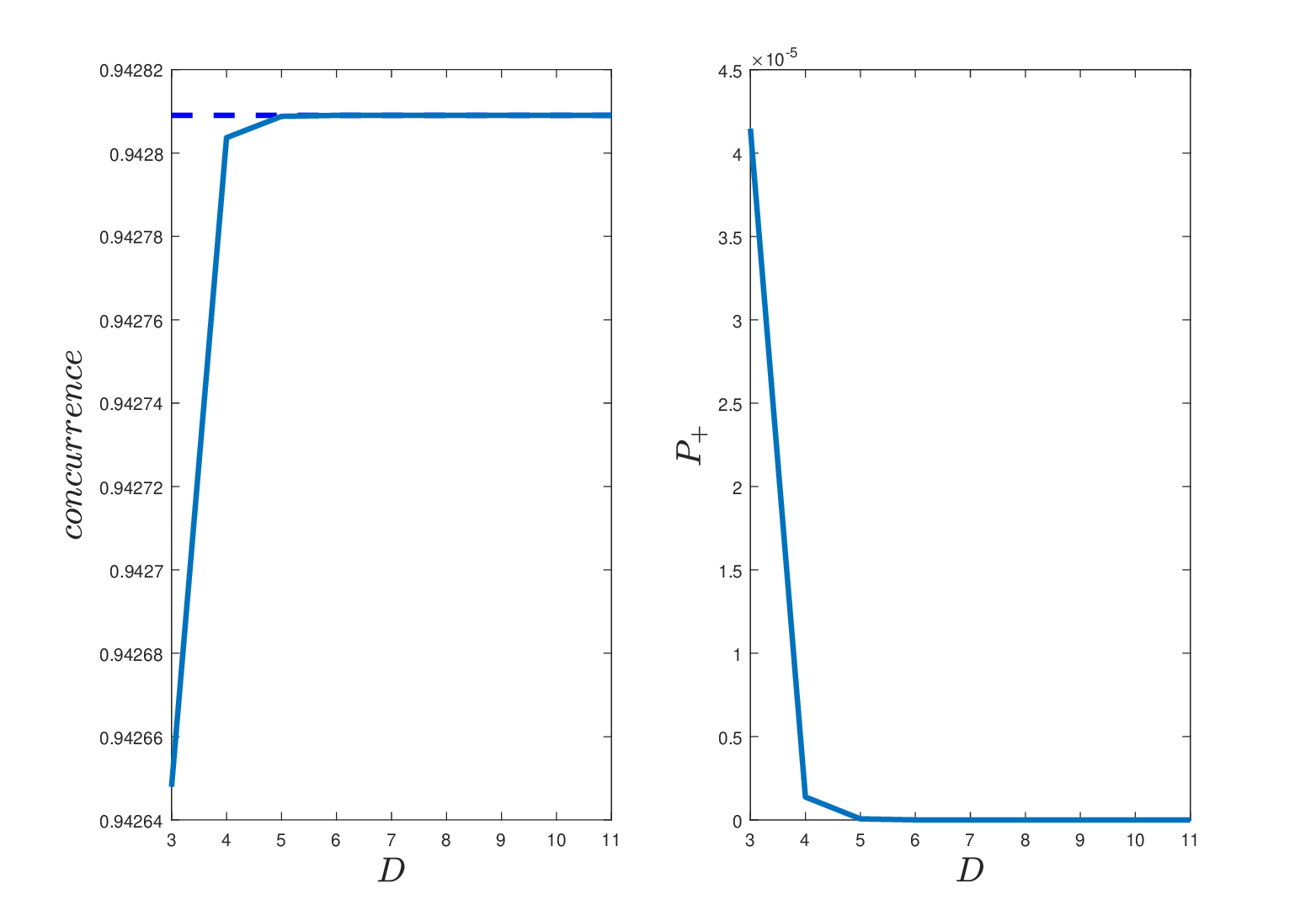} \caption{(Color online) The
left panel sows the concurrence of two static detectors in the 'long
interaction time' case with respect to the dimension $D$ of the spacetime. The
dashed blue line represents the initial entanglement. The solid blue line
represents the entanglement after interaction with the vacuum. The right panel
shows the corresponding transition probability from the ground state to the
excited state. Parameters are the same as in Fig. 3 except $a=0$.}%
\label{Fig6}%
\end{figure}

\begin{figure}[ptb]
\centering
\includegraphics[width=0.7\columnwidth]{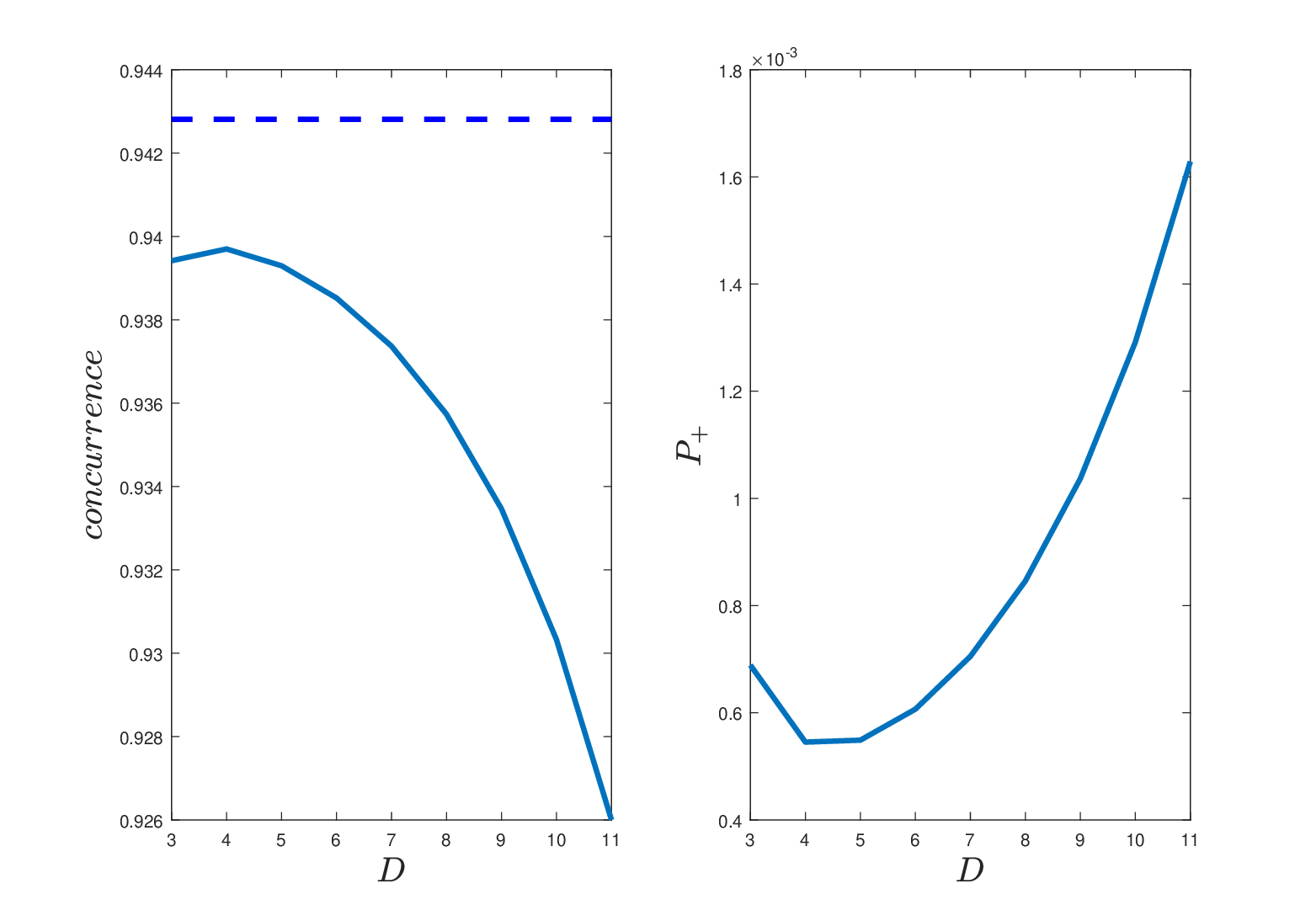} \caption{(Color online) The
left panel shows the concurrence of two static detectors in the 'short
interaction time' case with respect to the dimension $D$ of the spacetime. The
right panel shows the corresponding transition probability from the ground
state to the excited state. Parameters are the same as in Fig. 3 except $a=0$
and $\sigma= 0.2$.}%
\label{Fig7}%
\end{figure}

\section{Conclusion}

In this paper, we revisit the entanglement change caused by the acceleration.
We extend such a phenomenon to arbitrary spacetime dimensions and find that in
some specific conditions, the lost entanglement could be recovered completely.
We investigate the case of two uniformly accelerated detectors and found that
the lost entanglement could be recovered in higher dimensional spacetime when
the acceleration is small ($a\lesssim c$), but the entanglement recovery rate
is lower compared with the limit case of zero acceleration; when the
acceleration is large ($a\gtrsim c$), the lost entanglement could not be
recovered because the thermal effect caused by the Unruh effect spoil the
entanglement recovery process. We also study the entanglement change caused by
the Unruh and anti-Unruh effects, we found that the anti-Unruh effect can only
maintain for a range of acceleration and then will change into the Unruh
effect, and the anti-Unruh effect can just recover part of the lost
entanglement. By analytical methods, we also found that the anti-Unruh effect
can only appear in two-dimensional spacetime and the exact condition is
$\sigma\Omega<\frac{1}{\sqrt{2}}$. Finally, we examined the limit case of zero
acceleration, and find that the entanglement recovery is related to the the
detector-field interaction time scale $\sigma$. When the detector-field
interaction time scale $\sigma$ is long enough compared with the energy gap
$\Omega$, the lost entanglement could be recovered completely; when the
timescale is short compared with the energy gap, the lost entanglement may
recover part first, and then continue to lose with increasing spacetime
dimensionality $D$; when the timescale is very short, the entanglement will
lose more in higher dimensional spacetime compared with the lower dimensional
case. That is because the structure of higher dimensional spacetime is more
complicated so the time taken to reveal the lost entanglement will be
prolonged as the dimension increases.

\section{Acknowledgments}

This work is supported by National Natural Science Foundation of China (NSFC)
with Grant Nos. 12375057, 11947301, 12047502, and the Fundamental Research
Funds for the Central Universities, China University of Geosciences (Wuhan)
with No. G1323523064.

\section{Appendix: the calculation of transition probabilities}

We start with the general expressions of transition probability,%
\begin{equation}
P_{\pm s}=\int\int d\tau d\tau^{\prime}W(\tau-\tau^{\prime})e^{-\tau
^{2}/2\sigma^{2}}e^{-\tau^{\prime2}/2\sigma^{2}}e^{-i\Omega(\tau-\tau^{\prime
})}. \label{gep}%
\end{equation}
Making a variable substitution $\tau-\tau^{\prime}=\Delta\tau$ and $\tau
+\tau^{\prime}=\overline{\tau}$, Eq. (\ref{gep}) could be written as
\begin{align}
P_{\pm s}  &  =\frac{1}{2}\int\int d\Delta\tau d\overline{\tau}W(\Delta
\tau)e^{-\Delta\tau^{2}/4\sigma^{2}}e^{-\overline{\tau}^{2}/4\sigma^{2}}e^{\mp
i\Omega\Delta\tau}\nonumber\\
&  =\frac{1}{2}\int e^{-\overline{\tau}^{2}/4\sigma^{2}}d\overline{\tau}\int
W(\Delta\tau)e^{-\Delta\tau^{2}/4\sigma^{2}}e^{\mp i\Omega\Delta\tau}%
d\Delta\tau\nonumber\\
&  =\sqrt{\pi}\sigma\int d\Delta\tau W(\Delta\tau)e^{-\Delta\tau^{2}%
/4\sigma^{2}}e^{\mp i\Omega\Delta\tau}%
\end{align}
In the same fashion, $X_{\pm s}$ can be written as
\begin{equation}
X_{+s}=X_{-s}=X_{\pm s}=\sqrt{\pi}\sigma e^{-\sigma^{2}\Omega^{2}}\int
d\Delta\tau e^{-\Delta\tau^{2}/4/\sigma^{2}}W(\Delta\tau).
\end{equation}

According to the Sokhotski-Plemelj formula
\begin{equation}
\frac{1}{(x\pm i\epsilon)^{n}}=\frac{1}{x^{n}}\pm\frac{(-1)^{n}}{(n-1)!}%
i\pi\delta^{(n-1)}(x),
\end{equation}
the Wightman function in Eq. (\ref{ws}) could be written as
\begin{equation}
W(\Delta\tau)=\mathcal{C}(-1)^{-\frac{D-2}{2}}\frac{1}{\Delta\tau^{D-2}}%
-\frac{(-1^{D-2})}{(D-3)!}i\pi\delta^{(D-3)}(\Delta\tau). \label{ebc}%
\end{equation}

Substituting this equation (\ref{ebc}) into Eq. (\ref{PPP}), we can obtain
\begin{equation}
P_{\pm s}=\sqrt{\pi}\sigma\mathcal{C}(-1)^{-\frac{D-2}{2}}\int d\Delta\tau
e^{-\Delta\tau^{2}/4\sigma^{2}}e^{\mp i\Omega\Delta\tau}(\frac{1}{\Delta
\tau^{D-2}}-\frac{(-1)^{D-2}}{(D-3)!}i\pi\delta^{(D-3)}(\Delta\tau)).
\label{wdf}%
\end{equation}
We denote the two items in Eq. (\ref{wdf}) as $P_{1}$ and $P_{2}$ respectively.

Firstly, we will deduce a more explicit expression of $P_{1}$. When $D$ is
odd,
\begin{align}
P_{1}  &  =-2i\sqrt{\pi}\sigma\mathcal{C}(-1)^{-\frac{D-2}{2}}\int d\tau
e^{-t^{2}/4\sigma^{2}}\sin{\pm\Omega\tau}\frac{1}{\tau^{(D-2)}}\nonumber\\
&  =2^{2-D}\pi^{\frac{3-D}{2}}\sigma^{5-D}\Omega_{1}F_{1}[2-D/2,3/2,-\sigma
^{2}\Omega^{2}].
\end{align}
When $D$ is even,
\begin{align}
P_{1}  &  =2\sqrt{\pi}\sigma\mathcal{C}(-1)^{-\frac{D-2}{2}}\int d\tau
e^{-\tau^{2}/4\sigma^{2}}\cos{\Omega\tau}\frac{1}{\tau^{(D-2)}}\nonumber\\
&  =(2i)^{-D}\pi^{\frac{1-D}{2}}(D-1)\sigma^{4-D}\Gamma(\frac{1-D}{2}%
)\Gamma(\frac{D-2}{2})_{1}F_{1}[\frac{3-D}{2},\frac{1}{2},-\sigma^{2}%
\Omega^{2}].
\end{align}

Now, to get a more explicit expression of $P_{2}$ we need to use some
properties of distribution functions. The action of a distribution $f_{2}$ on
a test function $f_{1}$ is defined as
\begin{equation}
\langle f_{2},f_{1}\rangle:= \int_{-\infty}^{\infty} f_{2}(x)f_{1}(x)dx,
\end{equation}
This formula has a derivative relation
\begin{equation}
\langle\frac{df_{2}}{dx},f\rangle= -\langle f_{2},\frac{df_{1}}{dx}\rangle.
\end{equation}
which could be deduced directly by integral by part. Using this relation, we
can get
\begin{equation}
\langle\delta^{(D-3)}(x),f_{1}(x)\rangle= (-1)^{(D-3)} f_{1}^{(D-3)}(0).
\end{equation}
Applying this equation to $P_{2}$, we can obtain the expression
\begin{equation}
P_{2} = \pi^{3/2}\sigma i^{1-D} \mathcal{C} f^{(D-3)}(0).
\end{equation}
The expression of $X$ could be deduced in the same manner.

For the accelerated case, substituting the Wightman functions into Eq.
(\ref{wa}), one obtains
\begin{align}
P_{\pm a}  &  =(P_{\pm a}-P_{\pm s})+P_{\pm s}\nonumber\\
&  =\mathcal{C}\int d\tau e^{-\tau^{2}/4\sigma^{2}}e^{\mp i\Omega\tau}%
(\frac{a}{2i}\frac{1}{\sinh(\frac{a(\tau-i\epsilon)}{2})})^{D-2}%
-(-1)^{-\frac{D-2}{2}}\frac{1}{(\tau-i\epsilon)^{D-2}}+P_{\pm s}\nonumber\\
&  =\mathcal{C}\int d\tau e^{-\tau^{2}/4\sigma^{2}}e^{\mp i\Omega\tau}%
i^{-D}2^{1-D}a^{-3+D}\pi^{\frac{1-D}{2}}\sigma\Gamma(\frac{D}{2})e^{\frac
{\mp2i\Omega\tau}{a}-\frac{\tau^{2}}{a^{2}\sigma^{2}}}\frac{\sinh
{(\tau-i\epsilon)}^{D-2}-(\tau-i\epsilon)^{D-2}}{(\tau-i\epsilon)^{D-2}%
\sinh{(\tau-i\epsilon)}^{D-2}}+P_{\pm s}%
\end{align}
The first term could be considered as a \textquotedblleft
correction\textquotedblright\ induced by acceleration to the transition
probability in the limit of zero acceleration.

\end{document}